\documentclass[pre,twocolumn,showpacs,preprintnumbers,amsmath,amssymb]{revtex4}
\usepackage{amsfonts,amsmath}
\usepackage{graphicx}
\usepackage{dcolumn}

\newcommand {\br} {\langle}
 \newcommand {\ke} {\rangle}
\newcommand {\pd}{\partial}
\newcommand {\beq}{\begin{equation}}
\newcommand{\eeq}{\end{equation}}
\newcommand {\bseq}{\begin{subequations}}
\newcommand{\eseq}{\end{subequations}}
\newcommand{\bal}{\begin{align}}
\newcommand{\eal}{\end{align}}

\newcommand{\lm}{\lambda}

\newcommand{\bg}{\bar{g}}
\newcommand{\blm}{\bar{\lambda}}
\newcommand{\s}{\sigma}
\newcommand{\dl}{\delta}

\newcommand{\ep}{\varepsilon}

\newcommand{\dphi}{\phi^{\dagger}}

\begin{document}
\setlength{\unitlength}{1mm}

   \title{Vicious walks with long-range interactions}
  \author {Igor Goncharenko, Ajay Gopinathan }
  \affiliation{School of Natural Sciences, University of California, Merced, California, 95343, USA}
    \begin{abstract}The asymptotic behaviour of the survival or reunion probability of vicious walks  
  with short-range interactions is generally well studied. In many realistic processes, however, walks interact with a long ranged potential that decays in $d$ dimensions with distance $r$ as $r^{-d-\sigma}$. 
We employ methods of renormalized field theory to study the effect of such long range interactions. 
We calculate, for the first time, the exponents describing the decay of the survival probability  for all values of parameters $\s$ and $d$ to first order in the double  expansion in $\varepsilon=2-d$ and $\dl=2-d-\s$. We show that there are several regions in the $\s-d$ plane corresponding to different scalings for survival and reunion probabilities. Furthermore, we calculate the leading logarithmic corrections for the first time.

   \end{abstract}
  \maketitle
  
\section{Introduction}

Systems consisting of diffusing particles or random walks interacting by means of a long-range  potential  are non-equilibrium systems, which describe different phenomena in physics, chemistry and biology. From a physical perspective they are used to study   metastable supercooled liquids \cite{Supercool, Dean}, 
melting in type-II high-temperature superconductors \cite{Nelson},  electron transport in quasi-one-dimensional conductors \cite{Quasi1d} and carbon nanotubes \cite{Nanotube}.  From a chemical viewpoint the interest in these systems lies in the fact that some diffusion-controlled reactions processes rely on the diffusion of long-range interacting particles which react after they are closer than an effective capture distance. Some examples include radiolysis in liquids \cite{ParkDeem}, 
 electronic energy transfer reactions \cite{Klafter} and a large variety of chemical reactions in amorphous media \cite{RDreview}. From a biological viewpoint, the investigation of these systems is helpful in  understanding 
the dynamics of interacting populations in terms of predator-prey models \cite{Krap-Redner, Bray} and  
 membrane inclusions with curvature-mediated interactions \cite{Reynwar1, Reynwar2}.  

Vicious walks (VW) are a class of non-intersecting random walks, where the process is terminated upon the first encounter between walkers \cite{Fisher}. 
The fundamental physical quantity describing VW is the survival probability which is defined  as the  probability  that no pair of particles has   collided up to time $t$. Diffusing particles or walks that are not allowed to meet each other but otherwise remain free, we call pure VW.  The behavior  of pure VW is generally well-known. The survival probability for such a system has been computed in the framework of renormalization group theory in arbitrary spatial dimensions up to  two-loop order \cite{Cardy, Bhat1, Bhat2}. These approximations have been confirmed by exact results available in one dimension from the solution of the boundary problem of the Fokker-Plank equation \cite{Krap-Redner, Bray}, using matrix model formalism \cite{Katori} and Bethe ansatz technique \cite{Derrida}. On the other hand the effect of long range interactions has been extensively investigated in many-body problems.
It has been shown that the existence of  long-range disorder leads to a rich phase diagram with interesting crossover effects \cite{Halp, Bla, Prud}. If the   potential is Coulomb-like ($\sim r^{-1-\s}$) then systems in one dimension 
 behave similar to a one-dimensional version of a Wigner crystal \cite{Wigncrist}
for $\s<0$ and similar to  a Luttinger liquid for $\s\ge 0$ \cite{Mor-Zab}. If the potential is logarithmic then in the long-time limit the dynamics of  particles 
are described by non-intersecting paths \cite{Hinrichsen,Katori}.
The generalization of VW that includes the effect of long range interactions has not attracted much attention in the literature. Up to our knowledge there was one attempt to study long-range VW \cite{Bhat3}. Here the authors  considered the case of a  long-range potential decaying as $gr^{-\s-d}$, where $g$  is a coupling constant. It was shown by applying the Wilson momentum shell renormalization group that only one of the critical exponents characterize long-range VW.
For a specific value of $\s$ ($\s=2-d$) they show that the exponent $\gamma$, which determines the decay of the asymptotic survival probability with time, is given by the expression:
\beq\label{sp_bhat}
\gamma= \frac{p(p-1)}{4}u_1,
\eeq
 where $p$ the number of VW in the system, $u_1=(\ep/2+[(\ep/2)^2 +g]^{1/2})$ and $\ep=2-d$. There are limitations to the above approach. First, it is restricted to a single form of the potential ($\sim r^{-2}$) and systems such as membrane inclusions and chemical reactions have different power-law potentials. Second, it considers  identical walkers but one would like to have results if the diffusion constant of all walkers are different. Finally it is not convenient to compute higher-loop corrections using the Wilson formalism.

In this paper we reconsider the problem of long-range VW using methods of Callan-Symanzyk renormalized field theory in
conjunction with an expansion in $\varepsilon=2-d$ and $\dl=2-d-\s$. We note that it is more convenient to compute logarithmic and higher loop corrections by using this method. We derive the asymptotics of the survival and reunion probability for all values of the parameters $(\s,d)$ for the first time.  

\begin{table}
\vspace{0.3cm}
\caption{\label{tab:table1}%
One-loop survival probability of $p$ sets of particles with $n_j$ particles in each set large-time asymptotic at different regions of the $\s-d$ plane. We refer to Figure 3 for specific value of $\s$ and $d$ in each region.}
\begin{ruledtabular}
\begin{tabular}{ccc}
Region & Survival probability & \\
\hline
 I & $t^{-(d-2)/2}+t^{-(d+\s-2)/2}$ & \\
 II & $t^{-\frac{1}{2}\sum_{ij}n_in_j\ep}$ &   \\
 III & $t^{-\frac{u_1}{2}\sum_{ij}n_in_j (1+\dl/2\log t)} $& \\
 IV & $t^{-(d-2)/2}$ &  \\
 V, $d=2$ & $t^{-\frac{\sqrt{g_0}}{2}\sum_{ij}n_in_j (1+\dl/2\log t)} $ & \\
 VI,  $\s=2-d$ & $t^{-\frac{u_1}{2}\sum_{ij}n_in_j}$ \footnote{$u_1$ is defined by the formula (\ref{sp_bhat}).} & \\
 
\end{tabular}
\end{ruledtabular}
\end{table}

In this paper we will show that there are several regions in $\s-d$  plane in which we have different behavior of the critical exponent. Our results are summarized in  Table I. We note that results on the line $\s+d=2$ have been obtained before \cite{Bhat3}. Regions I and IV correspond to Gaussian or mean-field behavior (see Figure 3). In region II we found that the system reproduces pure VW. Logarithmic corrections in region III and at the short-range upper critical dimension $d=2$  have been obtained as series expansion in $\dl=2-\s-d$.

The remainder of this paper is organized as follows: 
Section~\ref{sec:model} reviews the field theoretic formulation of long range VW and describes Feynman rules and dimensionalities of various quantities.  In section~\ref{sec:LR} we derive the value of all fixed points and study their stability. Section~\ref{sec:results} presents results for the critical exponents and logarithmic corrections of various dynamical observables. Section~\ref{sec:concl} contains our concluding remarks. In Appendix A we give the details of the computation of some integrals that appear in Section~\ref{sec:LR}.

\section{Modelling VW with long-range interations}
\label{sec:model}

As the  starting point of the description of our model we consider $p$ sets of diffusing particles or random walks with $n_i$ particles in each set $i=1\dots p$, with a pairwise intraset interaction which includes a local or short-range part and a non-local or long-range tail. The local part determines the vicious nature of walks: if two walks belonging to the different sets are brought close to each other, both are  annihilated. Walks belonging to the same set are supposed to be independent. At $t=0$ all particles start in the vicinity of the origin. We are interested in the survival and reunion probabilities of walks at time $t>0$.

A continuum description of a system of $N$ Brownian particles $X_i$ with two-body interactions is simplified by the coarse-graining procedure in which a large number of microscopic degrees of freedom are averaged out. Their influence is simply modelled as a Gaussian noise-term in the Langevin equations. A convenient starting point for the description of the stochastic dynamics is the path-integral formalism. Then the system under consideration is modeled by the classical action 
\beq\label{langLR} S= \int\limits_0^{+\infty}dt  \left(\sum\limits_{i=1}^N \dot X_i^2/(2D_i) + \sum\limits_{i<j}V(X_i-X_j) \right)\eeq
where $t$ is (imaginary-)time, $X_i (t)$ is the $d$-dimensional vector denoting the position of $i$th particle at time $t$. 
$D_i$ is an $i$th particle diffusion coefficient. The path-integral representation of the probability density function for the particle displacements from their original positions is given by the functional ${\cal Z}=\int{\cal D}X\exp[-S]$. 
 The survival probability is defined as the expectation value 
 \beq\label{sp}P(t)=\br\prod_{i,j}[1-\delta(X_i(t)-X_j(t))]\ke \eeq 
 with respect to the functional ${\cal Z}$. It is computed in the framework of usual perturbation theory and will be a sum of integrals over internal degrees of freedom. It is more convenient to perform these integrations in Fourier space. To do this we would need the Fourier transform of the interaction potential $V(r)$. We note that it is comprised of a short-range part of the form $V_0(r) = \lm\delta(r)$ and a long-range part which decays with the distance $r$ as a power law, $V_l(r)= g r^{-d-\sigma}$. The Fourier transform of the latter is divergent if $\s\geq0$.
We introduce the cut-off parameter $a$ to regularize the singularity 
$V_l(r)=g(r^{2}+a^{2})^{-(d+\sigma)/2}$.
Fourier transformation of this function is given by the expression \beq\label{Vft}V_l(q)= g\frac{\pi^{d/2}2^{\s}}{\Gamma(\frac{d+\s}{2})}(q/a)^{\sigma/2}K_{\sigma/2}(aq),\eeq 
where $K_{\sigma}$ is the modified Bessel-function with index $\sigma$. Small $a$ expansion of (\ref{Vft}) at leading order yields  

\beq V_l(q)  \sim g\begin{cases} 
q^{\sigma}, & \mbox{if} \quad\s\ne 0\\
\log (aq) , & \mbox{if } \quad\s=0 \\
\end{cases} \eeq
where we used the property $K_{-\s}(x)=K_{\s}(x)$ of the Bessel function. The non-universal coefficient coming from the Taylor expansion can be absorbed by the appropriate renormalization of the constant $g$. Special cases when $\s$ is even gives  logarithmic behavior. Effectively it does not change  our results. So we focus on the typical term $ q^{\sigma}$.

The second quantized version of the action (\ref{langLR}) can be constructed using  standard methods 
\cite{Doi, Peliti}. The generalization of the action to the long-range interacting case is also known \cite{Mahan, Fetter}. The result is

\begin{widetext} \beq\label{hamil}
S (\phi_i,\dphi_i) = \int dtd^dx \{\sum_i [ \dphi_i\pd_{t} \phi_i + D_i \nabla\dphi_i\nabla \phi_i]\}  +\int dtd^dxd^dy \sum\limits_{i<j}  \dphi_i(t,x)\phi_i(t,x)V_{ij}(x-y)\dphi_j(t,y)\phi_j(t,y) \;. \eeq
\end{widetext}

The first term describes the evolution of free random walks with diffusion constants $D_i$. 
The potential is 
\beq V_{ij}(x-y) = \lm_{ij}\delta(x-y) + g_{ij} V(x-y),\eeq 
and we refer to $\lm_{ij}, g_{ij} $ as short-range and long-range coupling constants respectively. 

A dynamic response functional  associated with the action  (\ref{hamil}) is
\beq\label{Z}
\mathcal{Z} = \int \mathcal{D}\phi\mathcal{D}\dphi e^{-S(\phi_i,\dphi_i) }
\eeq
where $\phi_i\left(x,t\right)$ is the complex scalar field.  After the quantization we may treat $\dphi_i\left( x,t\right) $ as the creation operator which creates a particle of sort $i$ at point $x$ at time $t$. 
Having the dynamic response functional, correlation functions can be computed as functional averages (path integrals) of monomials of $\phi$ and $\dphi$ with the weight $\exp\left\{  -S(\phi,\dphi)\right\}$. 

\begin{figure}
\includegraphics[scale=0.35]{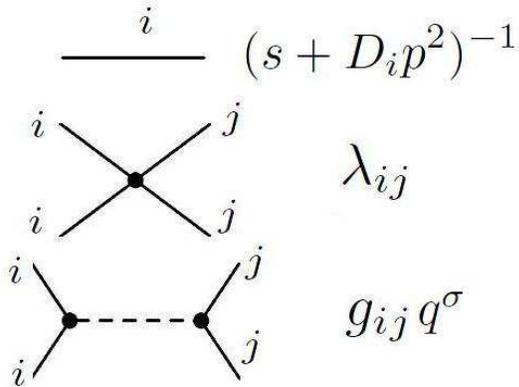}
\caption{\label{fig:frules} Feynman rules for the theory (\ref{hamil}). Notice that both $\lm$ and $g$ vertices appear with differenet $i$ and $j$ indices and that $g$ has momentum dependence.}
\end{figure}

\begin{figure*}
\includegraphics[scale=0.6]{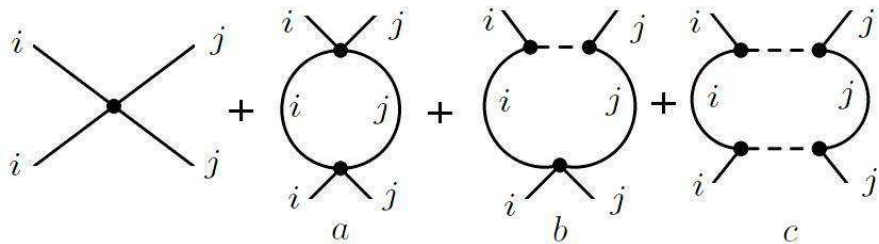}
\caption{\label{fig:fdiag} One-loop Feynman diagrams contributing to $\lm_{Rij}$}
\end{figure*}

As a first step towards the renormalization group  analysis of this model, we discuss the dimensions of various quantities in (\ref{hamil}) expressed in terms of momentum: \beq[t]=p^{-2}\quad [\phi]=p^d \quad [\lm]=p^{2-d} \quad [g]=p^{2-d-\s}.\eeq 
The naive dimension of the coupling constant $g$ allows us to identify the upper critical dimension $d_{c}(\s)=2-d-\sigma$.
For $\s>0$, the short-range term naively dominates the long-range term and we expect to have the behavior of the system similar to the  case of pure VW. 
We will reserve the symbol $\varepsilon$ ($\varepsilon=2-d$) to denote deviations from the short-range critical dimension $d_c=2$,   and  $\dl$ ($\dl=2-d-\s$) for the deviations from the long-range critical dimension $d_{c}(\s)$.
If $\s=0$ then the critical dimension of the long-range part coincides with the short-range part and we have the non-trivial correction to the asymptotic behavior due to long-range interactions.
 This boundary separates mean-field or Gaussian behavior  from  long-range behavior. 
For $\s<0$ the long-range term  dominates the short-range term and we expect to have non-trivial corrections to the behavior of the system. 

Now we consider diagrammatic representation elements of model (\ref{hamil}).
In zero-loop approximation the vertex 4-point function takes a simpler form after  Laplace-Fourier transformation: 
\beq \Gamma^{(2,2)}_{ij}(s,p) = V_{ij}(p_1+p_2) \delta(\sum\limits_k p_k).\eeq
The same transformation applied to the bare propagator yields:
\beq\label{prop} \Gamma^{(1,1)}_{j}(s,p) = (s+D_ip^2)^{-1} 
\eeq
We note that there are no vertices in (\ref{hamil}) that produce diagrams which dress the propagator, implying there is no field renormalization. As a consequence the bare propagator (\ref{prop}) is the full propagator for the theory. Feynman rules are summarized in Figure 1. There are two vertices in the theory: one is a short-range $\lm$-vertex and another is a long-range momentum dependent $g$-vertex. Each external line of the vertex corresponds to a functionally independent field. The propagator is formed by contracting appropriate lines from different vertices. We recall the propagator is the correlation function of $\phi_i$ and $\dphi_i$ fields only. 

Physical observables are  computed with the help of correlation functions. The probability that $p$ sets of particles with $n_i$ particles in each set  start at the proximity of the origin and finish at $x_{i,\alpha_i}$ ($i$ index  enumerates different sets and $\alpha_i$ index enumerates particles in set $i$) without intersecting each other can be obtained by generalizing eqn (\ref{sp}). 
In the field theoretical formulation, this probability becomes the following correlation function:
\begin{equation}
\label{sp-G}
 G(t)
=\int\prod_{i=1}^p\prod_{\alpha_i=1}^{n_i}d^dx_{i,\alpha_i}
\langle\phi_i(t,x_{i,\alpha_i})(\dphi_i(0,0))^{n_i}\rangle,
\end{equation}
In the Feynman representation it is the vertex with $2N$ ($N=\sum_j n_j$) external lines. In the first order of the perturbation theory one needs to contract these lines with corresponding lines of the vertices in Figure 1. Since there are many independent fields in the correlation function (\ref{sp-G}) this operation can be done in many ways. It yields a combinatorial factor, $n_in_j$, in  front of each diagram, which is the number of ways of constructing a loop from the $n_i$ lines of type $i$ and $n_j$ lines of type $j$ on the one hand and one line of type $i$ and one line of type $j$ on the other hand. From the next section we will see that  the survival probability scales as $G(t)\sim t^{-\gamma}$, where $\gamma$ is the critical exponent. If all walks are free, $\gamma=0$. In the presence of interactions we expect $\gamma$ to be a universal quantity that does not depend on the intensity of the short-range interaction $\lm_{ij}$. It is convenient to introduce the so called truncated correlation function which is obtained from (\ref{sp-G}) by factoring out external lines:
\beq\label{tcf} \Gamma(t) = G(t)/(\Gamma^{(1,1)})^{2N}\eeq

 Another physical observable, the reunion probability, is defined as the probability that $p$ sets of particles with $n_i$ particles in each set  start at the proximity of the origin and without colliding into each other finish at the proximity of some point at time $t$: 
\beq\label{rp}
 R(t)
=\int d^dx \prod_{i=1}^p \langle\phi_i(t,x)^{n_i}(\dphi_i(0,0))^{n_i}\rangle,
\eeq
In the Feynman representation  it is depicted as the watermelon diagram with  $2N$ stripes.
We note that if the theory is free this expression is the product of free propagators and at the large-time limit the return probability  scales  as $R_{\cal O}(t) \sim t^{-(N-1)d/2}$.
If interactions are taken into account it becomes $R(t) \sim t^{-(N-1)d/2 -2\gamma}$, where $\gamma$ is survival probability exponent. The reason that it enters with the factor 2 is the following. If we cut a watermelon diagram of the reunion probability correlation function in the middle then it produces  two vertex diagrams with $2N$ external lines of the survival probability correlation function. As a result the reunion probability is the product of two survival probabilities. It remains true in all orders of perturbation theory. For a rigorous proof we refer to \cite{Bhat2}.

\section{The Renormalization of observables}
\label{sec:LR}

  \begin{figure}[b]
\includegraphics[scale=0.25]{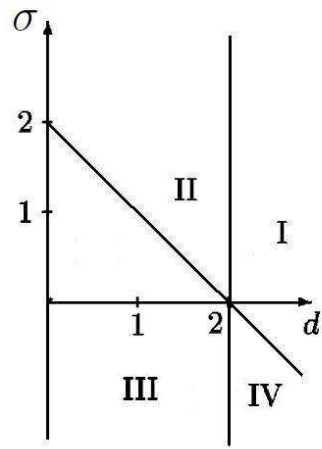}
\caption{\label{fig:sigmad} The critical behavior of vicious walks with long-range interactions in the different regions of the $(\s,d)$ plane. Region I and IV correspond to the mean field short-range behavior, in region II will be critical short-range behavior, region III is the long-range behavior. The lines $d=2$ and $\s+d=2$ represent regions V and VI respectively.}
\end{figure}

While computing correlation functions like (\ref{sp-G})  perturbatively one faces divergent integrals when $d=d_c$.
The convenient scheme developed for dealing with these divergences follows Callan-Symanzik renormalization-group analysis \cite{Zinn, Amit}. Within this scheme we start with the bare correlation function 
$G(t;\lm,g)$, where $\lm =\{\lm_{ij}\}$, and $g=\{g_{ij}\}$ denote the set of bare short-range and long-range coupling constants. 
In the renormalized theory it becomes $G_{R}(t;\lm_R,g_R,\mu)$. From  dimensional analysis it follows that 
\beq\label{diman} G_{R}(t;\lm_R,g_R,\mu) = G_{R}(t\mu;\lm_R,g_R),\eeq
where $\mu$ is the renormalization scale. The scale invariance leads to the expression 
\beq\label{scaleinv} G_{R}(t;\lm_R,g_R,\mu) = Z(\lm_R, g_R, \mu)G(t;\lm,g).\eeq
Here functions $Z$ are chosen in such a way that   $G_{R}(t,\lm_R,g_R,l)$ remains finite when the cut-off is removed at each order  in a  series expansion of $\lm_R$, $g_R$, $\ep$ and $\dl$. From the fact that $G(t,\lm,g)$ does not depend on the renormalization scale $\mu$ we get the Callan-Symanzik equation 
\beq\label{cseq} \left(\mu\frac{\pd}{\pd \mu} + \beta_g \frac{\pd}{\pd g} + \beta_u \frac{\pd}{\pd u} - \gamma \right) G_{R}=0, \eeq 
where the $\beta$-functions are defined by 
\beq\label{beta} \beta_{\lm} (\lm_R, g_R)= \mu  \frac{\pd}{\pd \mu} \lm_R\qquad  \beta_{g} (\lm_R, g_R)=\mu \frac{\pd}{\pd \mu} g_R \eeq and the function  $\gamma$ by \beq\label{gamma}\gamma(\lm_R, g_R) = \mu \frac{\pd}{\pd \mu} \ln Z.\eeq
The renormalization group functions are understood as the expansion in double series of coupling constants $\lm$ and $g$ and deviations from the critical dimension $\ep$ and $\dl$. We take $\dl = O(\ep)$.
The coefficient $Z(\lm_R, g_R, \mu)$ is fixed by the normalization conditions. It is more convenient to impose these conditions on the Laplace transform of the truncated correlation function (\ref{tcf}). One sets the following condition then 
\beq\label{norm}  \Gamma_{R}(\mu) =1,\eeq
when $s=\mu$. 
We note that the same  multiplicative renormalization factor $Z$ yields $\Gamma$ finite. From this fact one can infer that 
\beq\label{scalGamma}\Gamma(\mu;\lm,g) = Z(\mu;\lm,g)^{-1}.\eeq
If we express unrenormalized couplings in terms of renormalized ones (\ref{scalGamma}) we will obtain the  equation for finding $Z$ explicitly.

The equation (\ref{cseq}) can be solved by the method of characteristics. Within this method we let couplings depend on the scale which is parametrized by $\mu(x)=x\mu $. Here $x$ is introduced as a parametrization variable of the RG flow and is not to be confused with position. Henceforth $x$ will refer to this parametrization variable. We introduce running couplings $\bar \lm(x)$ and $\bar g(x)$. They satisfy the equations \beq x\frac{d}{dx}\bar g(x)=\beta_g(\bar \lm(x),\bar g(x))\quad x\frac{d}{dx}\bar \lm(x)=\beta_{\lm}(\bar \lm(x),\bar g(x)).\eeq The renormalized value should be defined by the initial conditions $\bar \lm(1)=\lm_R$ and $\bar g(1)=g_R$. the solution of the equation  is then
\beq\label{solRG} G_{R}(t)
= e^{\int\limits_1^{\mu t} \gamma(\bar \lm(x), \bar g(x))dx/x}G_{R}(\mu^{-1};\bar\lm(\mu t),\bar g(\mu t),\mu) \eeq

Next we calculate the first-order contribution to the   renormalized vertices.
The $\lm$-vertex  is   renormalized by the set of diagrams that are shown in Figure 2. We notice that there are no diagrams producing the momentum dependent $g$-vertex in the theory (\ref{hamil}). 
This statement is the corollary of the fact that only independent fields of power one enter into the expression of the vertex and there are no higher powers of fields. Also we keep in mind that the renormalized couplings are defined by the value of the vertex function  taken at zero external momenta. It produces the following expression:
\beq\label{Run}\begin{cases} 
\lm_{Rij} &= \lm_{ij} -\frac{1}{2} (\lm_{ij}^2 I_1 + 2\lm_{ij} g_{ij}I_2 + g_{ij}^2I_3)\\
g_{Rij} &= g_{ij} \\
\end{cases} \eeq
 where $I_k=I_k(\s;D_i,D_j)$ are one-loop integrals corresponding to the diagrams $a$, $b$, $c$ in the Figure 2 respectively. Using the Feynman rules we can explicitly write them down:
 \beq\label{int} I_k = \int \frac{d^dq}{(2\pi)^d} \frac{q^{(k-1)\s}}{2s+(D_i+D_j)q^2}, \quad k=1,2,3. \eeq
  We will use dimensional regularization procedure to compute these integrals. The details of the computation are summarized in  Appendix A.  We note that integrals will diverge logarithmically at different values of the spatial dimension $d$. For this reason it leads to different critical behavior in different regions of the $\s-d$ plane (see Figure 3). These regions correspond to four possibilities for $\ep=2-d$ and $\dl=2-d-\s$ to be positive or negative. Only if $\dl=O(\ep)$ or, in other words, if both $\ep$ and $\dl$ are infinitesimally small but the ratio $\ep/\dl$ is finite we expect non-zero fixed points of the renormalization group flow. Similar approximation have been used before \cite{Halp} but for different models with long-range disorder. It allows us to follow the standard procedure of deriving the $\beta$-functions which consists of two steps.  
  
  First, we express  unrenormalized couplings in terms of the renormalized.   For the short-range  coupling constant $\lm$ it can be done by solving the quadratic equation in (\ref{Run}). Expanding the square root and keeping terms up to the second order we infer that 
\beq\label{unR}\begin{cases} 
\lm_{ij} &= \lm_{Rij} +\frac{1}{2} (\lm_{Rij}^2 \frac{a_d}{\ep} + 2\lm_{Rij} g_{Rij}\frac{b_d}{\dl} + g_{Rij}^2\frac{c_d}{2\dl-\ep})\\
g_{ij} &= g_{Rij} \\
\end{cases} \eeq
where $a_d$, $b_d$ and $c_d$coefficients have been found explicitly in  Appendix A. Now we introduce dimensionless renormalized couplings 
\beq\label{dless} \bg_{Rij} = a_d(2s)^{-\dl/2} \quad \blm_{Rij} = b_d(2s)^{-\ep/2}.\eeq An important observation is that $c_da_d=b_d^2$ which can be verified by  explicit substitution (see Appendix A). Multiplying the first and second equation in (\ref{unR}) by the factors $a_d$ and $b_d$ respectively, and using redefinitions (\ref{dless}) we can condense all pre-factors in the right hand side of the equations into the dimensionless constants. 

Second, we differentiate equations (\ref{unR}) with respect to the  scaling parameter $\mu$. Using definitions (\ref{beta}) and the fact that bare couplings do not depend on the scale,   we derive 
\beq\label{betalmg}\begin{cases} 
\beta_{\lm,ij} &= -\ep \bar\lm_{Rij} + (\bar\lm_{Rij}+ \bar g_{Rij})^2\\
\beta_{g,ij} &= -\dl \bar g_{Rij} \\
\end{cases} \eeq
where the right hand side is understood as the leading contribution to the $\beta$-functions from the double expansions in 
$\lm,g $ and $\ep,\dl$. 
 From (\ref{betalmg}) we see that it is convenient to introduce new coupling constants $u_{Rij}= \bar \lm_{Rij} +\bar g_{Rij}$. After this step the  renormalization group equations read

 \beq\label{flow}\begin{cases} 
\beta_{u,ij} &= -\ep u_{Rij} + u_{Rij}^2 -g_{Rij} \\
\beta_{g,ij} &= -\dl g_{Rij} \\
\end{cases} \eeq
We note that in the last equations $g$ coupling constant  has been redefined $\s \bar g_{Rij}\to g_{Rij}$.

Fixed points are zeros of the $\beta$-functions. If $\dl \neq 0$ then the last equation in (\ref{flow}) is zero only when $g_*=0$. Then the first equation has two solutions $u=0$ and $u=\ep$. If $\dl=0$ then $g$ plays the role of a parameter and the fixed points are determined by the roots of the quadratic equation
\beq 0=-\ep u + u^2 -g\eeq
which are real if $g\ge-(\ep/2)^2$ and we find
\beq u_{1,2} = \ep/2 \pm \sqrt{(\ep/2)^2 + g}.\eeq
All fixed points are listed in the Table II. The stability of these fixed points is determined by the matrix of partial derivatives 
\beq\label{stab} \beta_* = - 
 \left( \begin{array}{cc}
\pd\beta_u/\pd u & \pd\beta_u/\pd g \\
\pd\beta_g/\pd u & \pd\beta_g/\pd g  
 \end{array} \right)_{u=u_*, g=g_*} \eeq
Eigenvalues are listed in the Table \ref{tab:fp}. The Gaussian fixed point is stable in all directions for $\ep<0$ and $\dl<0$ which corresponds to region I in  Figure 3. In this region we find both  short-range(pure VW) and  long-range  mean-field behavior depending on the sign of $\s$. 
On the contrary, for $\ep>0$ and $\dl>0$ we find that the Gaussian fixed point is unstable(irrelevant) in all directions and the  short-range (pure VW) fixed point is stable(relevant) only in $u$-direction. It means that  long-range interactions will play a leading  role. This region corresponds to region  III in  Figure 3. Next for $\ep>0$ and $\dl<0$ we find that the short-range (pure VW) fixed point is stable in all directions. It means that the system is insensitive to the long-range tail.  This region corresponds to region  II in  Figure 3. Finally for $\ep<0$ and $\dl>0$ we find that the short-range (pure VW) fixed point is unstable in all directions and the system will be described by  mean-field at long time.

\begin{table}[b]
\caption{\label{tab:fp}
Fixed points for flow equations (\ref{flow}) and the corresponding eigenvalues $(\lm_1,\lm_2)$ of the stability matrix (\ref{stab}). We note that $u_1$ and $u_2$ are values of the }
\begin{ruledtabular}
\begin{tabular}{ccc}
Fixed point & $(u_*, g_*)$ & $(\lm_1, \lm_2)$ \\
\hline
 Gaussian & $(0,0)$ & $(\ep,\dl)$ \\
Pure VW  & $(\ep,0)$ & $(-\ep,\dl)$ \\
 LR stable  & $(u_{1},0)$ & $(-\sqrt{\ep^2-4g},0)$ \\
 LR unstable & $(u_{2},0)$ & $(\sqrt{\ep^2-4g},0)$
 \end{tabular}
\end{ruledtabular}
\end{table}

\section{Calculation of  critical exponents and discussion}
\label{sec:results}

Here we describe our  method of computing critical exponents. It is based on the formula (\ref{scalGamma}) from the previous section. First, we obtain the leading divergent part of the correlation function. 
The renormalized correlation function depends on the scale $\mu$ but it appears in all formulas in combination with time: $\mu t$. 
Second, since we have found the  bare coupling constant as a function of renormalized (dressed) couplings we express correlation function in terms of dressed couplings. 
Finally using the normalization condition (\ref{norm}) and the definition (\ref{gamma}) we differentiate $Z$ with respect to $\mu \pd/\pd\mu$ to obtain the exponent $\gamma$. The poles should cancel after this operation. 

 In section 2 it was explained  that the truncated correlation function in the one-loop approximation is given by the formula
\beq\label{oneloop} \Gamma(t;\lm,g)= 1- \sum_{i,j} n_i n_j\left(\lm_{ij} I_1 +g_{ij}I_2\right).\eeq
Here integrals are the same as in (\ref{int}). 

We start our analysis with the region I. Notice that truncated correlation function $\Gamma(t)$ and survival probability $G(t)$ have similar large time behavior. We use large momentum cut-off to compute  integrals $I_1$ and $I_2$ as in formula (\ref{mfint}) in Appendix A. The renormalization of coupling constants is trivial in this case.
Therefore the leading contribution to the survival probability is given by 
\beq G(t)\sim t^{(2-d)/2}+g_0t^{(2-d-\s)/2},\eeq 
where $g_0$ is non-universal coefficient and we will not need its exact value. We notice that if $\s>0$ the second term will decay faster than the first term and in the long-time limit it will produce the same behavior as mean-field pure VW. On the other hand if $\s<0$
the first term will decay faster and long-range interactions will play a leading role.  Many authors observed similar behavior in various systems with long-range defects \cite{Halp, Bla, Prud}. Intuitively if potential falls fast with distance than the system effectively represent system with short-range potential where particle interact when they are close to each other. 

Region IV exhibits similar behavior. Now the integral $I_2$ is computed with the help of the dimensional regularization (\ref{intres}) and the integral $I_1$ remains the same. From the fact (\ref{diman}) one can infer that
the survival probability scales as 
\beq G(t)\sim t^{(2-d)/2}.\eeq
Short-range behavior dominates because  the running coupling constant will flow towards the  Gaussian fixed point  at long time limit which is the only stable fixed in this region. This result is exact regardless the number of loops one takes into account.

In Region II the computation is as follows. 
\beq\label{rtwo} \ln Z = \sum n_in_j\left( \lm_{ij}\frac{a_d}{\ep} + g_{ij}t^{(2-d-\s)/2}\right),\eeq
so plugging the result from (\ref{ad}) to (\ref{rtwo}) we obtain at the fixed point $(\lm_*=\ep, g=0)$
\beq\label{gtwo} \gamma = -\frac{1}{2}\sum n_in_j\ep \eeq
And we reproduce the pure VW behavior.  This result is the reflection of the fact that the renormalization-group trajectories run away to stable pure VW fixed point. It is with agreement with the results obtained by Katori in \cite{Katori} for $d=1$, and the logarithmic intraset particle interactions. The irrelevance of the long-range interaction in lower dimensions is a typical phenomenon observed in a various out of equilibrium interacting
particle systems.

We now consider regions III, V and VI.  Integrals in (\ref{oneloop}) are computed via dimensional regularization. Taking the inverse of (\ref{oneloop}) and then logarithm one can obtain at the leading order:  
\beq\label{lnZ} \log Z = \sum n_in_j\left(\lm_{ij}\frac{a_d}{\ep} + g_{ij}\frac{b_d}{\dl}\right)\eeq
where  $a_d$ and $b_d$ are defined in Appendix A in (\ref{ad}) and (\ref{bd}).
We note that after taking the derivative the poles in (\ref{lnZ}) will cancel in the limit of $\dl=O(\ep)$. Also one recalls the expansion (\ref{unR}) and the redefinitions in (\ref{dless}). Using (\ref{gamma}) we show that the expression for the function $\gamma$ which determines critical exponent takes the form
\beq \gamma =- \frac{1}{2}\sum_{ij} n_in_j u_R \eeq
 Evaluated at the stable fixed point  $(u_1=\ep/2+ \sqrt{(\ep/2)^2 +g}$ it gives the following result:
\beq\label{gthree} \gamma = -\frac{1}{2}\sum_{ij}  n_in_j u_1,  \eeq
and the survival probability scales as  $G(t)\sim t^{\gamma}$.

We will now find the logarithmic corrections to this scaling law. The running coupling constant can be found from the flow equation (\ref{flow}):  $\bar{g}(x) = e^{-\dl x} g$. In the  case $\dl,\ep=0$ (the intersection of regions V and VI) the flow equation for  $\bar u(x)$ is 
\beq\label{flowu} x\frac{d\bar u(x)}{dx} = -\bar{u}^2(x) +g \eeq and the solution is
\beq\label{tanh} \bar{u}(x) = \sqrt{g} \tanh (\sqrt{g} \log x +\phi_0)\sim \sqrt{g} \tanh (\sqrt{g} \log x),\eeq
where $\phi_0$ is the initial condition and we do not need its exact form. After plugging this expression into the (\ref{solRG}) we infer 
\beq\label{gammaint} \int\limits_1^{\mu t} \gamma(\bar u, \bar g)\frac{dx}{x}  \sim \log (\cosh (\sqrt{g} \log \mu t))\eeq
 Thus the survival probability is
\beq\label{dnulld2} G(t) \sim \cosh (\sqrt{g} \log t)^{-\frac{1}{2}\sum n_in_j } \eeq
In the limit of large time $\cosh (\sqrt{g} \log t)\sim t^{\sqrt{g}} $ implying $gamma =- \frac{1}{2}\sum_{ij} n_in_j \sqrt{g}$ which is consistent with equation (\ref{gthree}). 
For negative coupling constant $g<0$ the solution in (\ref{tanh}) becomes
\beq\label{tan} \bar{u}(x) \sim -\sqrt{|g|} \tan (\sqrt{|g|} \log x)\eeq
The integral (\ref{gammaint}) is divergent if $t>\exp(\pi/2\sqrt{|g|})$ which leads to the result that the survival probability is zero beyond this time. For smaller times one has  $G(t) \sim \cos (\sqrt{|g|} \log t)^{-\frac{1}{2}\sum n_in_j }$. Thus, upto one-loop order approximation, It implies that if walks are attracted to each other then all of them will annihilate at some finite time. This might be a signature of faster than power law decay and we expect to have corrections to this behavior at higher loop approximation.

Next we consider the case when $\ep=0$ and $\dl\ne 0$ but $\dl$ remains small i.e. region V. The flow equation for the $\bar u(x)$ is 
\beq\label{flowug} x\frac{d\bar u(x)}{dx} = -\bar{u}^2(x) +g x^{-\dl}\eeq and the solution can be found by the method of perturbation. 
Up to the first order
\beq \bar u(x) = \sqrt{g}\tanh (\sqrt{g} \log x)+\dl\sqrt{g}\log(x)\tanh (\sqrt{g} \log x)\eeq
After plugging this expression into eqn (\ref{solRG}) we infer 
\beq \int\limits_1^{\mu t} \gamma(\bar u, \bar g)\frac{dx}{x}  \sim -\frac{1}{2}\sum n_in_j\left(\log(t^{\sqrt{g}}) +\frac{1}{2}\dl\sqrt{g} \log^2 (t)\right) \eeq
Therefore we have the correction to the survival probability in the form
  \beq G\sim  t^{-\frac{1}{2}\sum n_in_j \sqrt{g} (1 + \delta/2(\log t))}\eeq

Now we extend our analysis to the case when $\ep>0$, corresponding to regions III and VI. The evolution of the coupling constant is \beq x\frac{d}{dx}\bar u(x) = \ep\bar u-\bar u^2 +g x^{\dl}\eeq
We choose the ansatz in the form $\bar u(x) = u_0(x)+\dl v(x)$. For $\dl=0$ (i.e. region VI) the equation for  $u_0(x)$ reads 
\beq \label{u0ex} x\frac{d}{dx} u_0(x) = \ep u_0- u_0^2 +g  \eeq
and we reproduce the result (\ref{gthree}). We now extend to the case where $\ep,\dl>0$ (region III). Here we will need the exact solution to (\ref{u0ex}) to find the corrections:
\beq  u_0(x) = \frac{Cx^{u_1-u_2} u_1 +u_2}{1+Cx^{u_1-u_2}}, \eeq where $C =(u_R-u_2)/(u_1-u_R)$. The logarithmic correction follows from the form of the perturbation. The equation for $v(x)$ is 
\beq x\frac{d}{dx}v(x) = \ep v-2u_0v -g \log x \eeq
 The solution can be found explicitly as a combination of hypergeometric functions. 
  In the most interesting case, $\ep=1$ ($d=1$) the hypergeometric functions are degenerate and become linear functions. Corrections to the integral then read
\beq \int\limits_1^{\mu t} \gamma dx/x\sim \frac{1}{2}\dl u_1 \log^2( t)
+\log( t)  ( t)^{u_1-u_2}) \eeq
In the limit of large time only the first term contributes to the exponent and the survival probability scales as 

\beq G\sim  t^{-\frac{1}{2}\sum n_in_j u_1 (1 + \delta/2\log t)}\eeq

\section{Conclusion}
\label{sec:concl}
In summary, we studied long-range vicious walks using the methods of Callan-Symanzik renormalized field theory.
 Our work confirms the previously known RG fixed point structure including their stability regions.
 We calculated 
the critical exponents for all values of $\s$ and $d$ to first order in $\ep$ expansion and to all orders in $\dl$ expansion, which have hitherto been known only for $d+\s=2$. Our results indicate that, depending on the exact values of $d$ and $\s$, the system can be dominated by either short range (pure VW) or long range behaviors.
In addition, we calculated the leading logarithmic corrections for several dynamical observables that are typically measured in simulations.

We hope that our work stimulates further interest in long-range vicious walks. It would be interesting to see further simulation results for the critical exponents for $d>1$ and for logarithmic corrections. Also, it would be interesting to have analytical and numerical results for other universal quantities such as scaling functions and amplitudes.

\section{Acknowledgments}
  AG would like to acknowledge UC Merced start-up funds and a James S. McDonnell Foundation Award for Studying Complex Systems.

\section*{Appendix A}

Effective four-point function (one-particle irreducible, 1PI) that appeared in (\ref{Run}) is composed of usual short-range  and new momentum dependent vertices. This gives rise to integrals (\ref{int}). The first integral $\mu=1$ has been evaluated in \cite{Cardy} by using  alpha representation $1/(q^2+s) = \int_0^{+\infty} d\alpha e^{i(q^2+s)\alpha}$ and the result is \beq I_1=K_d (2s)^{-\ep/2}\Gamma(\ep/2).\eeq We notice that since there is no angular dependence one can perform $d-1$ integrations and one will be left with one dimensional integral. To compute this integral we use the formula \cite{GR}:

\beq \int\limits_{0}^{+\infty} dx \frac{x^{\nu-1}}{P+Qx^2} = \frac{1}{2P} \left(\frac{P}{Q}\right)^{\nu/2} \Gamma\left(\frac{\nu}{2}\right)\Gamma\left(1-\frac{\nu}{2}\right)\eeq
We see that in our case $P=s$, $Q=(D_i+D_j)$ and $\nu =d+(\mu-1)\s$. This immediately gives the result:

\begin{align}  I_{\mu} =&  \frac{K_d}{2} \left(\frac{1}{(D_i+D_j)}\right)^{\frac{d+(\mu-1)\s}{2}} s^{\frac{d+(\mu-1)\s}{2}-1} \times
\nonumber\\
&\times\Gamma\left(\frac{d+(\mu-1)\s}{2}\right)\Gamma\left(1-\frac{d+(\mu-1)\s}{2}\right)
\,, \label{intres}%
\end{align}
where $K_d = 2^{d-1}\pi^{-d/2}\Gamma^{-1}(d/2)$ is the surface area of $d$-dimensional unit sphere. 

It is convenient to define \beq\label{ad} a_d = \frac{K_d}{2} \left(\frac{2}{(D_i+D_j)}\right)^{d/2} (2s)^{-\ep/2}\eeq
\beq\label{bd} b_d = \frac{K_d}{2} \left(\frac{2}{(D_i+D_j)}\right)^{(d+\s)/2} (2s)^{-\dl/2}\eeq
\beq\label{cd} c_d = \frac{K_d}{2} \left(\frac{2}{(D_i+D_j)}\right)^{(d+2\s)/2} (2s)^{-(2\dl-\ep)/2}\eeq
So integral $I_{\mu}$ in the limit of $\dl=O(\ep)$ can be written as:
\beq I_1=\frac{a_d}{\ep},\quad I_2=\frac{b_d}{\dl},\quad I_3=\frac{c_d}{2\dl-\ep}.\eeq
We used an expansion $\Gamma(\ep/2) \sim 2/\ep$ for small $\ep$.
An important property of coefficients (\ref{ad}) - (\ref{cd}) is that \beq c_da_d=b^2_d,\eeq
which can be verified by direct substitution.

Now we compute mean field integrals:

\beq\label{mfint} I_{\mu} = \int d^dq dt q^{d+\s}\exp(-t(D_i+D_j)q^2)\sim t^{-(d+\s-2)/2}, \eeq
where we assumed that the large momentum cut-off is imposed and corresponding coupling constants have been renormalized. The non-universal coefficient  is not important.

\end{document}